\documentclass[10pt]{iopart}
\usepackage{graphicx}
\usepackage{latexsym}
\usepackage{dcolumn}
\usepackage{bm}
\usepackage{color}
\usepackage{amssymb}

\newcommand{\be}{\begin{equation}}
\newcommand{\ee}{\end{equation}}
\newcommand{\bea}{\begin{eqnarray}}
\newcommand{\eea}{\end{eqnarray}}
\newcommand{\bse}{\begin{subequations}}
\newcommand{\ese}{\end{subequations}}

\begin{document}

\paper[]{Universality in the time correlations of the long-range 1d Ising model}
\date{\today}

\author{Federico Corberi$^{1}$, Eugenio Lippiello$^{2}$, Paolo Politi$^{3,4}$}

\address{$^{1}$ Dipartimento di Fisica ``E.~R. Caianiello'', and INFN, Gruppo Collegato di Salerno, 
and CNISM, Unit\`a di Salerno,Universit\`a  di Salerno, via Giovanni Paolo II 132, 84084 Fisciano (SA), Italy.}
\address{$^{2}$ Dipartimento di Matematica e Fisica, Universit\`a della Campania,
Viale Lincoln 5, 81100, Caserta, Italy}
\address{$^{3}$ Istituto dei Sistemi Complessi, Consiglio Nazionale
delle Ricerche, via Madonna del Piano 10, I-50019 Sesto Fiorentino, Italy}
\address{$^{4}$ Istituto Nazionale di Fisica Nucleare, Sezione di Firenze, via G. Sansone 1 I-50019, Sesto Fiorentino, Italy}
\ead{corberi@sa.infn.it, eugenio.lippiello@unicampania.it, paolo.politi@isc.cnr.it}

\begin{abstract}
The equilibrium and nonequilibrium properties of ferromagnetic systems may be affected by the long-range 
nature of the coupling interaction.
Here we study the phase separation process of a one-dimensional Ising model in the presence of a power-law decaying 
coupling, $J(r)=1/r^{1+\sigma}$ with $\sigma >0$, and we focus on
the two-time autocorrelation function $C(t,t_w)=\langle s_i(t) s_i(t_w)\rangle$. 
We find that it obeys the scaling form $C(t,t_w)=f(L(t_w)/L(t))$, where $L(t)$ is the typical domain size at time $t$, and 
where $f(x)$ can only be of two types.
For $\sigma>1$, when domain walls diffuse freely, $f(x)$  falls in the nearest-neighbour (nn) universality class.
Conversely, for $\sigma \le 1$, when domain walls dynamics is driven, $f(x)$ displays a new universal behavior.
In particular, the so-called Fisher-Huse exponent, which characterizes the asymptotic behavior  of $f(x)\simeq x^{-\lambda}$
for $x\gg 1$, is $\lambda=1$ in the nn universality class ($\sigma > 1$) and $\lambda=1/2$ for $\sigma \le 1$.
\end{abstract}
\noindent{\bf Keywords:} Coarsening processes; Correlation functions; Kinetic Ising models; Numerical simulations

\submitto{Journal of Statistical Mechanics: theory and experiment}

\maketitle

\section{Introduction} \label{intro}

Universality means that seemingly different phenomena may be characterized by similar quantities.
This concept has proved extremely useful when classifying equilibrium phase transitions~\cite{Huang}, which can
be characterized by a limited number of critical exponents, depending on very
general features of the physical system under study: the space dimension, the symmetry of the order
parameter, the short/long range character of the interaction, the absence/presence of quenched disorder.
Nonequilibrium phenomena have a so large variety of behaviors that
general classifications are more problematic even in specific domains~\cite{libro}, like absorbing phase transitions or kinetic roughening 
phenomena.

Phase ordering~\cite{Bray94}, the topic of this paper, is another nonequilibrium process whose comprehension 
can profit from universal concepts like symmetries and conservation laws
and the Ising model has played a special role in understanding this field.
In practice, it is a matter of studying the relaxation to equilibrium after a temperature quench from the disordered
phase ($T_i>T_c$, here $T_i=\infty$) to the ordered one ($T_f\equiv T < T_c$, or very low $T$ if $T_c=0$).
Two large universality classes are well known for the pure, short-range Ising model, according to the conservation or not
of the order parameter during the relaxation process. If we introduce the dynamical exponent $z$ to characterize
the coarsening process~\cite{CRP}, i.e. the increase over time of the average domain size, $L(t) \simeq t^{1/z}$,
we have $z=2$ for nonconserved order parameter and $z=3$ for conserved one, regardless of the space dimension.

If long-range interactions are present, their effects on equilibrium properties are well studied~\cite{review_long_range,Mukamel2009}.
In particular, it is known that sufficiently strong long-range interactions lead to loss
of additivity and to nonequivalence of statistical ensembles. We will confine our study
to the case where such phenomena do not occur (weak long-range interactions).
In real systems, there are many examples of (strong and weak) long-range, power-law interactions.
They vary from gravitational to magnetic (dipole and RKKY) interactions, from
elastic and hydrodynamic to Coulomb ones. A recent and detailed discussion about applications
can be found in the book~\cite{book_long_range}.

As for the nonequilibrium phase ordering process, recent analyses~\cite{EPL,ourreview} have revealed
unexpected results for the asymptotic conserved dynamics and for the transient regimes, if long-range 
interactions are present.
More precisely, with  the algebraic spin-spin coupling $J(r) =1/r^{1+\sigma}$ 
the results for the asymptotic nonconserved dynamics confirm the expected scenarios as given by the 
continuum theory
of Bray and Rutenberg~\cite{BrayRut94,RutBray94}: if $\sigma >1$ long-range interactions are irrelevant 
(see next Section for details), i.e. $T_c=0$, 
and one has $z=2$. Instead, for $\sigma \le 1$ they are relevant, namely $T_c>0$,
and it is  $z=1+\sigma$. 
In particular, we remark that the dynamical exponent keeps continuous through the value $\sigma=1$,
which is the lower limit of $\sigma$ for an integrable coupling, i.e. $\sum_{r=1}^{\infty} J(r) < \infty$.

Relaxation dynamics is not uniquely characterized by the coarsening exponent $1/z$, 
hence it is also interesting to study other features of the domain structure, e.g. the size distribution 
of domains~\cite{AreBrayCugSic07,SicSarAreBrayCug09},
or the behaviour of two-time correlation functions~\cite{Furukawa89,CorCugYos11,PhysRevE.93.052105}.
The exact solution of the one dimensional kinetic Ising model, due to the late Roy Glauber~\cite{Glauber1963},
tells that the autocorrelation spin function takes a scaling form $\langle s_i(t) s_i(t_w) \rangle = f(L(t_w)/L(t))$,
with $f(x) \simeq x^{-\lambda}$
for $x\gg 1$ ($t\gg t_w$), where $\lambda=1$ is the so called Fisher-Huse exponent~\cite{Fisher88}.

In this paper we are going to study how the scaling function $f(x)$ and the value of $\lambda$ 
are affected by the presence of long-range
interactions, finding the following results: (i)~$\lambda=1$ if long-range interactions are integrable,
i.e. $\sigma >1$, while $\lambda=1/2$ if they are not integrable ($0<\sigma \le 1$). Therefore, there is
a discontinuity at $\sigma=1$. (ii)~We propose an interpretation of these two classes through a qualitative
feature of domain wall (DW) dynamics.
It can be shown~\cite{ourreview} that DW asymptotically diffuse freely if $z=2$ 
(i.e. when $\sigma >1$) and
that they are driven when $z=\sigma +1$ (for $\sigma \le 1$). Then we argue that 
the free/driven character of DW diffusion is the quality distinguishing between $\lambda=1$ and
$\lambda=1/2$. The value of $\lambda$ for $\sigma=1$ supports this picture. In fact, for $\sigma=1$
the drift gives the same coarsening law as the free diffusion case, because in this case $z=1+\sigma=2$, 
but the
asymmetric DW hopping makes $\lambda=1/2$, while $\lambda=1$ for $\sigma=1^+$.
(iii)~Not only the exponent $\lambda$, but also the scaling function $f(x)$ displays
only two universal behaviors, depending on $\sigma > 1$ or $\sigma \le 1$.

\section{Model and observable quantities} \label{model}

We consider the Ising model in one dimension whose Hamiltonian reads
\be
   {\cal H}=-\frac{1}{2}\sum _{i,j=1}^{N}J(|i-j|) s_is_j \, ,
   \label{ham}
\ee
where $s_i=\pm 1$ are binary spin variables,
$J(r)=r^{-(1+\sigma)}$ ($\sigma >0 $), and we use
periodic boundary conditions.
Let us mention that letting
$J(r)= \delta _{r,1}$ we recover the usual Ising model with nearest
neighbors (nn) interactions.

The equilibrium properties of the model~(\ref{ham}) are well known.
For $\sigma >1$ the system falls in the universality class of the nn Ising model,
hence $T_c=0$ and the magnetisation vanishes at any finite
temperature~\cite{Peierls1934,Dyson1969}.
This is simply due to the fact that a domain wall has a finite energy cost
$E_{dw} = 4\sum_{r=1}^{\infty} J(r) < \infty$.
For $\sigma =1$ there is a Kosterlitz-Thouless phase
transition~\cite{Frohlich1982,Imbrie1988,Luijten2001} with a
discontinuity of the order parameter.
For $0<\sigma < 1$ there is a second-order phase transition at a finite critical
temperature $T_c$~\cite{Dyson1969}. In particular, in the range $\frac{1}{2}<\sigma <1$
the critical exponents depend continuously on $\sigma $~\cite{Fisher1972critical} whereas
for $0<\sigma \le \frac{1}{2}$ fluctuations are negligible and the
transition is in the mean-field universality class~\cite{Mukamel2009}.
For $\sigma \le 0$ additivity and extensivity are lost. We will not
consider here this strong long-range case~\cite{review_long_range}.

The model~(\ref{ham}) can be endowed with a dynamics by flipping single
spins and therefore does not conserve the magnetisation, namely we are
considering a model with nonconserved order parameter.

After a quench from a high temperature (that in the following we will
consider for simplicity to be infinite) to a low temperature $T$
the evolution of the system is characterised by a coarsening process
where spins order in a domain structure and the typical size $L(t)$ of
such domains grows in time. Operatively, in a numerical simulation $L(t)$ can be computed as
the inverse of the density of misaligned spins.

This kinetic process has been investigated
in~\cite{ourreview} where it was shown that the system behaves differently
in different time domains and for different $\sigma$.
The asymptotic $\sigma-$dependence can be easily rationalized as follows.
In the nn model, domain walls are free to diffuse so a pair of DW at distance $L$
takes a time $L^2$ to annihilate, hence the value $z=2$.
If we add long-range interactions, the same pair of DW has an energy $U(L)$ which
is obtained summing all the couplings $J(r)$ between a spin inside the domain
and a spin outside the domain. In practice, we must integrate twice $J(r)$.
Upon deriving $U(L)$ one obtains the force $F(L)$ acting between DW,
and in an overdamped picture of DW motion their velocity $v(L)$ is proportional to that force.
Finally, we obtain
\be
F(L) \sim \int_L^\infty dr J(r) \sim \frac{1}{L^\sigma} \propto v(L) , 
\ee
and since $v(L)=L/t$ 
the annihilation time scales as $t\sim L^{\sigma +1}$.
The deterministic drift acts when it is faster than symmetric hopping, $1+\sigma < 2$.
Therefore, for $\sigma < 1$, $z=1+\sigma$ and DW move asymmetrically while for $\sigma > 1$,
$z=2$ and DW move through a symmetric hopping. The former regime will be called
Rutenberg-Bray (RB) regime~\cite{BrayRut94,RutBray94}, the latter is the diffusive regime.

The above considerations are valid at finite temperature because at $T=0$ the Glauber flipping rate
(\ref{Glauber}) makes DWs move deterministically (one towards the other) with a constant velocity,
which gives rise to a ballistic regime, characterized by $z=1$.

Simulations and more rigorous calculations~\cite{ourreview} provide the results summarized in Table~\ref{table},
which must be understood with time $t$ increasing from left to right:
at short times there is always the ballistic regime ($z=1$); at finite $T$ this regime is replaced
by the RB regime ($z=\sigma+1$) when $L\gg L_1$; finally, if $\sigma >1$ we have the diffusive regime ($z=2$) when $L\gg L_2$. The quantities $L_1,L_2$ above are characteristic lengths where crossovers
between the various regimes discussed insofar take place.
It is worth noting that for $\sigma=1$ both RB and diffusive mechanisms give the same annihilation time ($t\sim L$)
but DW dynamics is actually asymmetric, because $L_2(\sigma=1^+)=\infty$.

\begin{table}
  \centering
  \small
      \begin{tabular}{|c|c|c|c|c|}
\hline
& & & & \\
ballistic & $L_1$ & Rutenberg-Bray (RB) & $L_2$ & diffusive \\
& & (asymptotic for $\sigma\le 1$) & & (only for $\sigma > 1$) \\
\hline 
& & & & \\
$t$ & $\left( \frac{2}{\sigma} \beta  \right)^{\frac{1}{\sigma}}$ & $t^{\frac{1}{\sigma +1}}$ & 
$\left( \frac{2}{\sigma } \beta \right)^{\frac{1}{\sigma -1}}$ & $t^{\frac{1}{2}}$ \\
& & & & \\
\hline
\end{tabular}
$$ ------------ \mbox{~~~~~time~~~~~} ------------> $$
\caption{
Summary of the different coarsening regimes as explained in the main text
and as discussed in details in Ref.~\cite{ourreview}.
The short time regime is always ballistic, while the asymptotic one 
depends on $\sigma$: it is diffusive ($z=2$) for $\sigma >1$ and it is RB
($z=\sigma+1$) for $\sigma \le 1$.}
\label{table}
\end{table}

Notice also that for $\sigma >1$ the final state is disordered at any finite
quench temperature. This means that coarsening is eventually interrupted
and equilibration is achieved in a final time even in the thermodynamic limit.
However, since this occurs when $L(t)\simeq \xi (T)$, where $\xi (T)$ is the
equilibrium coherence length that diverges as $T\to 0$, at low
temperatures the coarsening stage lasts for a huge time.
In the following we will always consider times much smaller than the
equilibration one. On the contrary, for $\sigma \le 1$ equilibration sets
in as due to a finite-size effect and can be postponed at will moving
towards the thermodynamic limit.

In this paper the observable we focus on is the autocorrelation function
\be
C(t,t_w)=\frac{1}{N}\sum _{i=1}^N\langle s_i(t) s_i(t_w)\rangle,
\label{defauto}
\ee
where $t_w<t$ and we take the average $\langle\cdots\rangle$ over the random initial condition 
$\{s_i(0)\}$ and over the thermal histories.
Notice that the subtraction of 
the disconnected term $\langle s_i(t)\rangle \langle s_i(t_w)\rangle$
is unessential since $\langle s_i(t)\rangle=0$ at any time in the coarsening stage.
The average over $i$ in Eq.~(\ref{defauto}) is taken only to improve the statistics,
since the average $\langle\cdots\rangle$ does not depend on $i$.

In the case with nearest neighbor (nn) interactions the autocorrelation function can be computed 
exactly~\cite{Bray_1990,Prados_1997}. In the case of a quench to $T=0$, for $t_w$
larger than a microscopic time $t_0$, and also for $t-t_w>t_0$,
one has a scaling form
\be
C(t,t_w)=f\left (\frac{L(t)}{L(t_w)}\right ),
\label{scalC}
\ee
with $L(t)\sim t^{\frac{1}{2}}$, and
\be
f(x)=f_{nn}(x)\equiv \frac{2}{\pi} \arcsin \sqrt {\frac{2}{1+x^2}}.
\label{analf}
\ee  
For quenches to $T>0$ the same form~(\ref{scalC},\ref{analf}) is obeyed in the coarsening stage
before equilibration takes place.
Notice that, from Eq.~(\ref{analf}), one has
\be
f(x)\sim x^{-\lambda} \quad \mbox{for}\,\,\,\, x\gg1,
\label{deflambda}
\ee
with the Fisher-Huse exponent $\lambda =1$.

In the system with space decaying interactions, while the dynamical scaling
form~(\ref{scalC}) is still expected, the scaling function $f(x)$
cannot be analytically computed and, to our knowledge,
has never been studied numerically. As we will see,
an algebraic decay as in Eq.~(\ref{deflambda}) is present also in this
case, but with an exponent $\lambda $ whose numerical value can be different
from the case with nn. Indeed, we have anticipated in Sec.~\ref{intro} that the presence of long-range interactions may
determine a switch from $\lambda=1$ to $\lambda=1/2$.

With regard to the variability of the Fisher-Huse exponent,
let us mention that $\lambda $ is expected to
be larger than $\lambda _{inf}=1/2$ and smaller than $\lambda _{sup}=1$.
Indeed, the lower bound $\lambda _{inf}=d/2$
of this exponent was found in arbitrary dimension in~\cite{Yeung96}
using arguments  
based on the properties of the structure factor. The same lower bound,
together with the upper bound $\lambda_{sup}= d$, was established in \cite{Fisher88}
using scaling arguments originally developed for spin glasses.

\section{Numerical simulations} \label{simul}

We adopt a fast simulation protocol where flips of spins in the bulk are forbidden and only  spins at the interface can flip. 
The dynamics is then mapped to the displacement of domain walls which can move towards the right or towards the left with a probability 
given by  transition rates of the Glauber type,
\be
w=\frac {1}{1+e^{\beta \Delta E}},
\label{Glauber}
\ee
where $\Delta E$ is the energy change associated to the spin reversal
and $\beta =(k_BT)^{-1}$ is the inverse temperature; in the
following we will set the Boltzman constant $k_B=1$.
$\Delta E$ is obtained from Eq.~(\ref{ham}) taking into account that, 
because of periodic boundary conditions, the distance between two sites $(i,j)$, with
$1\le i,j \le N$, is the minimum between $|j-i|$ and $(N-|j-i|)$.
Annihilation between two domain walls occur as soon as they reach the same position.

We have considered a sufficiently large systems size $N=4\times 10^6$ 
to avoid finite size effects. Indeed, for this choice of $N$, $N^{-1-\sigma}$ is so small that
the interaction between spins at distances larger than $N$ can be neglected
and we have explicitly verified that our results are $N-$independent. 
Furthermore, we have also checked that a sufficient number of domain walls $n_{DW}\gtrsim 10$ is  present in the system at all times.

Since the properties of the model are different in the two sectors
$\sigma >1$ and $\sigma \le 1$
we discuss results of numerical simulations in these two situations separately below.

\subsection{$\sigma >1$} \label{sg1}

\begin{figure}[h]
 \centering
\rotatebox{0}{\resizebox{.85\textwidth}{!}{\includegraphics{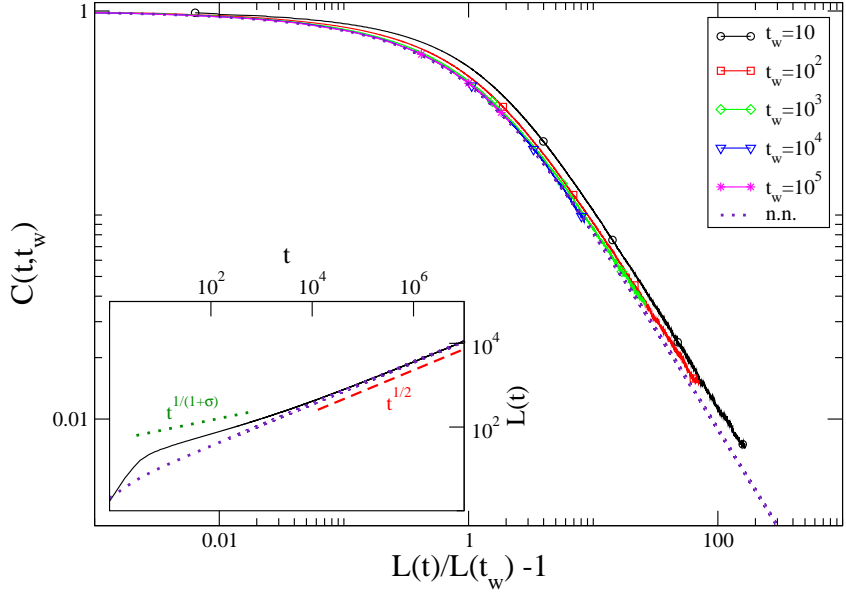}}}
\caption{Main: $C(t,t_w)$ is plotted against $L(t)/L(t_w)-1$ (the -1 subtraction is used to better show the
small $L(t)-L(t_w)$ behavior) for a quench to $T_f=10^{-2}$ with $\sigma =3$ 
and different values of $t_w$, as reported in the legend.
The dotted indigo curve is the scaling form~(\ref{analf}) of the nn case.
Inset: We plot $L(t)$. The dotted green line is the behavior
$t^{1/(1+\sigma)}$ expected in the RB regime and the dashed red line is
the form $t^{1/2}$ typical of the diffusive stage. The dotted indigo curve is the nn case. 
}
\label{fig_C_sigma_3}
\end{figure}

In this case, according to the discussion of the previous Section,
the asymptotic regime is characterized by an unbiased diffusion of the DW. 
The behavior of $L(t)$ for the model
with $\sigma =3$ quenched to $T=10^{-2}$ is plotted in the inset of
Fig.~\ref{fig_C_sigma_3}, hinting at the pre-asymptotic regime\footnote{%
The transient RB regime can be made more visible~\cite{ourreview} by tuning $T$ and $\sigma$
in order to make $L_2$ larger (see Table~\ref{table}). 
Here, however, we focus only on the asymptotic regime.}
where $L(t)\sim t^{1/(1+\sigma)}$
and showing the asymptotic diffusive regime, where $L(t)\sim t^{1/2}$.

Let us now discuss the behavior of the autocorrelation function, which is
plotted in the main part of the figure. In this case 
we expect to be in the same universality class of the
nn model. Hence $C(t,t_w)$ should behave, for sufficiently long times, as
in Eqs.~(\ref{scalC},\ref{analf}). Indeed one observes that when $t_w$ is
chosen in the asymptotic regime, the scaling form of $C(t,t_w)$ is indistinguishable from the
analytic solution $f_{nn}(x)$, plotted with a dotted indigo curve.
This is true already for $t_w=10^3$.
It is interesting to note that, for values of $t_w$ small enough to
belong to the preasymptotic stage ($t_w=10,10^2$),
although $f(L(t)/L(t_w))$ is displaced with respect to $f_{nn}(x)$,
it decays asymptotically as in Eq.~(\ref{deflambda}) and with
the same exponent $\lambda $ of the nn case, regardless of
the fact that $L(t_w)$ in this case is quite different from the one of
the nn case (see inset). This shows that the exponent $\lambda $ is
not changed by the kinetics around $t_w$, but is only determined by the
decorrelation mechanisms acting at $t\gg t_w$.
\vspace{1.2cm}

\subsection{$\sigma \le1$} \label{sl1}

\begin{figure}[h]
\vspace{1.5cm}
  \centering
 \rotatebox{0}{\resizebox{.85\textwidth}{!}{\includegraphics{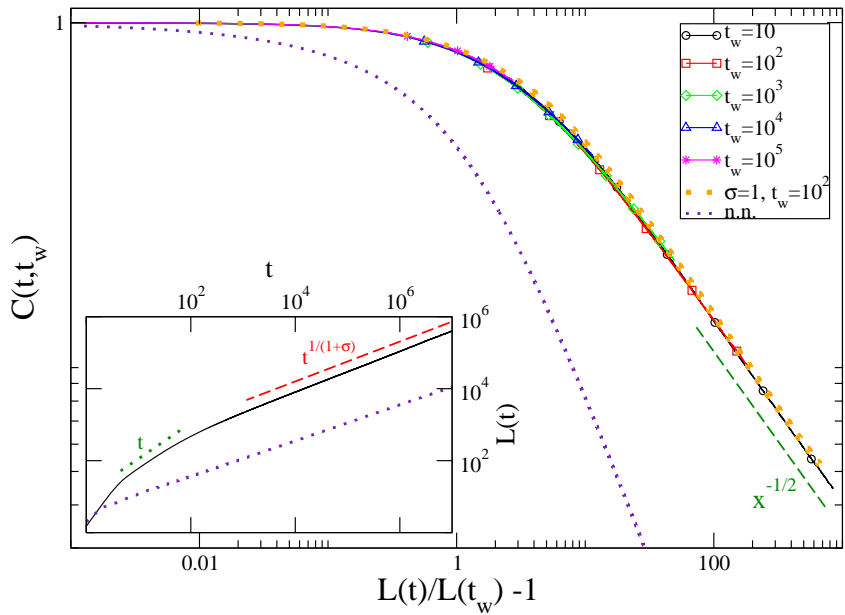}}}
\caption{Main: $C(t,t_w)$ is plotted against $L(t)/L(t_w)-1$ (the -1 subtraction is used to better show the
  small $L(t)-L(t_w)$ behavior) for a quench to $T=10^{-1}$ with $\sigma =0.8$
  or to $T=10^{-2}$ for $\sigma =1$ (dotted orange line)
and different values of $t_w$, as reported in the legend.
The dashed green line represents the behaviors $x^{-1/2}$. 
The dotted indigo curve is the scaling form~(\ref{analf}) of the nn case.
The two scaling functions differ for the full form of $f(x)$, Eq.~(5),
not only for the exponent $\lambda$, Eq.~(7).
Inset: We plot $L(t)$ for the quench to $T=10^{-1}$ with $\sigma =0.8$.
The dashed red line is the asymptotic RB behavior $L(t)\sim t^{1/(1+\sigma)}$,
while the dotted green one is the ballistic behavior $L(t)\sim t$.
The dotted indigo curve is the nn case.}
\label{fig_C_sigma_less_1}
\end{figure}

According to the discussion of Sec.~\ref{model}, for $\sigma \le 1$,
after the ballistic stage the system enters the RB regime with $z=1+\sigma$
which, in this case, is the asymptotic one.
For the particular case with $\sigma =0.8$ and for $T=10^{-1}$
both these regimes
can be observed, as shown in the inset of Fig.~\ref{fig_C_sigma_less_1}.
Notice also the marked different evolution,
at any time, of the long range model with respect to the nn one (dotted indigo line).

The behavior of the autocorrelation function is shown in the main part
of Fig.~\ref{fig_C_sigma_less_1}. 
Continuous curves with symbols correspond to $\sigma =0.8$ and $T=10^{-1}$,
for different choices of $t_w$ (see key). The smaller value of $t_w$
corresponds to the beginning of the ballistic regime while the larger values
belong to the asymptotic RB regime.
One observes that all the
curves exhibit a nice collapse when plotted against $L(t)/L(t_w)$,
meaning that the dynamical scaling form~(\ref{scalC}) is very well obeyed
also in this case. However the master curve $f$ is much different from the one
of the nn case (dotted indigo curve) and, in particular, the large
$L(t)/L(t_w)$ behavior is given by Eq.~(\ref{deflambda}) with
a value of $\lambda $ very well consistent with $\lambda =1/2$, largely different
from the one ($\lambda =1$) of the nn case.
 
Repeating the calculations for different values of $0<\sigma\le1$ and different values of $T$ 
we find the same pattern of behavior
with the same scaling function. For instance, data for $\sigma =1$ and
$T=10^{-2}$ are reported
as a heavy dotted orange curve in Fig.~\ref{fig_C_sigma_less_1}.
The above implies that the whole scaling function, not only the value of $\lambda$, is universal and 
independent both on temperature
and on the value of $\sigma$ (provided $0< \sigma \le 1$).

Let us notice that, although we know that the regime with $z=1+\sigma$ is the asymptotic one up to $\sigma =1$, 
with $\sigma =1$ one has $L(t)\sim t^{1/2}$, exactly as in
the diffusive stage, which is asymptotic for $\sigma $ strictly larger than
$\sigma =1$. 
Hence, by looking at $L(t)$ alone, one could not say if
one is in a diffusive regime with or without bias  of DWs. Instead, the scaling functions
of the autocorrelation function are markedly different in the two cases, as shown in Fig.~\ref{fig_C_sigma_less_1}. This quantity, therefore, is able
to distinguish unumbiguously between a kinetics with purely diffusive
interfaces and another where diffusion occurs with a drift.
 
\section{Conclusions}

The phase ordering dynamics following a temperature quench  can be characterized by
the time dependent spin-correlation function $\langle s_i(t)s_j(t_w)\rangle$, whose
special cases are the equal time  correlation 
 $C(r,t)=\langle s_i(t)s_{i+r}(t)\rangle $, which is traditionally considered in coarsening
 studies and the autocorrelation function considered in this paper, $C(t,t_w)=\langle s_i(t)s_i(t_w)\rangle$.

The spin-correlation functions are useful tools to discuss the issue of universality in 
phase-ordering kinetics.
In such a non-equilibrium setting, at variance with equilibrium, we don't know a priori what are 
the ``relevant" or ``irrelevant" couplings. It is worth noting that, in order to establish the
relevance-irrelevance of a parameter $\epsilon $ it is not sufficient to check if the large scale properties 
of  a given observable quantity ${\cal O}$ depend on $\epsilon $. This is because,
due to some hidden symmetry, ${\cal O}$ might not depend on $\epsilon $
even if this is relevant. In this case, observables which do not share the symmetries of ${\cal O}$
will possibly depend on $\epsilon$, thus informing us about its relevance.
An example of ``apparent irrelevance" is provided by the parameter $d$ in the context of growth kinetics with nn coupling:
while $d$ does not affect the dynamical exponent $z$, i.e. the coarsening law,
it changes the Fisher-Huse exponent $\lambda$, which increases with $d$~\cite{Fisher88,PhysRevB.44.9185}.

The picture emerging from the one-dimensional analysis~\cite{ourreview} of the exponent $z$ in the presence of 
algebraic $J(r)$ shows that the issue of universality in the coarsening-ordering process is simple, but not trivial:
in the nonconserved case long-range interactions are relevant for $\sigma <1$, similarly to the
equilibrium case, while in the conserved case they are unexpectedly irrelevant for any $\sigma >0$. 

In this paper we have discussed nonconserved dynamics, showing that the exponent $\lambda$ has a
discontinuity at $\sigma=1$, being equal to 1 (the value of the nn model) for $\sigma >1$ and
equal to 1/2 for $\sigma \le 1$. This sharp classification applies not only to the exponent $\lambda$,
but to the full scaling function $f(x)$.

The fact that $\lambda(\sigma)$ has a discontinuity in contrast to $z(\sigma)$, led us to think that
such different behaviors might be related to the breaking of a symmetry in the DW motion which
is irrelevant for equal time correlations, but not for two time correlations:
for $\sigma >1$ DW diffuse freely while for $\sigma\le 1$ long-range interactions are
asymptotically relevant and DW feel a drift towards the closest DW.
In order to have a further check on this hypotheses, besides the cases discussed above, we have 
carried out numerical simulations in a different asymptotic regime with biased motion of the DW:
the ballistic one in a quench to $T=0$. Performing an analysis (not shown here) analogous to the one 
displayed in Fig.~\ref{fig_C_sigma_less_1} we found $\lambda =1/2$ also in this case. 
This provides a further evidence to our conjecture.

We conclude by remarking that the two values of $\lambda$, $\lambda=1/2$ for $\sigma\le 1$ and
$\lambda=1$ for $\sigma > 1$, correspond to $\lambda_{min}$ and $\lambda_{max}$, respectively.
We don't know yet if this is a coincidence or if the switching from symmetric to biased DW diffusion
is the key ingredient to obtain $\lambda=\lambda_{min}$.
Such an interpretation, as well as a simple and physically oriented
derivation of the Fisher-Huse exponents would be very welcome.

\vskip 2cm



\end{document}